\documentclass[twocolumn,showkeys,aps,prb,showpacs]{revtex4-1}
\usepackage{graphicx}
\usepackage[CJKbookmarks,dvipdfm,colorlinks,linkcolor=blue,citecolor=blue]{hyperref}

\begin{document}

\title{Small compressive strain induced semiconductor-metal transition and tensile strain enhanced thermoelectric properties in  monolayer $\mathrm{PtTe_2}$}

\author{San-Dong Guo}
\affiliation{Department of Physics, School of Sciences, China University of Mining and
Technology, Xuzhou 221116, Jiangsu, China}
\begin{abstract}
 Biaxial strain  effects on electronic structures and  thermoelectric properties of   monolayer $\mathrm{PtTe_2}$ are investigated by using  generalized gradient approximation (GGA) plus spin-orbit coupling (SOC) for the electron part and GGA for
the phonon part. Calculated results show that small compressive strain (about -3\%) can induce semiconductor-to-metal transition, which can easily be achieved in experiment. The conduction bands convergence is observed for unstrained $\mathrm{PtTe_2}$, which can be removed by both compressive and tensile strains.  Tensile strain can give rise to valence bands convergence by changing  the position of  valence band maximum (VBM), which can induce enhanced Seebeck coefficient, being favorable for high power factor. It is found that tensile strain can also reduce lattice thermal conductivity, which  at the strain of 4\% can decrease by about 19\% compared to unstrained one  at room temperature. By considering tensile strain effects on $ZT_e$ and lattice thermal conductivity, tensile strain indeed can improve p-type  efficiency of thermoelectric conversion. Our results demonstrate the potential of  strain engineering in  $\mathrm{PtTe_2}$ for applications in  electronics and thermoelectricity.

\end{abstract}

\pacs{72.15.Jf, 71.20.-b, 71.70.Ej, 79.10.-n ~~~~~~~~~~~~~~~~~~~~~~~~~~~~~~~~~~~Email:guosd@cumt.edu.cn}

\maketitle

\section{Introduction}
Due to  potential application in electronic and energy conversion devices, two-dimensional (2D) materials have been widely investigated both in theory and experiment\cite{q1,q2,q3,q3-1}. Since the discovery of graphene,
great efforts have been made to explore 2D materials, such as hexagonal
boron nitride\cite{q4}, germanene, silicene\cite{q6}, phosphorene\cite{q7},
group IV-VI compounds\cite{q8}and transition metal
dichalcogenides\cite{q9}. The transition metal
dichalcogenides monolayer $\mathrm{MoS_2}$  possesses   intrinsic direct band gap of 1.9 eV in comparison with the gapless Graphene, and has been applied in field effect transistors, photovoltaics and photocatalysis\cite{q10,q11,q12}.
The thermoelectric properties related with $\mathrm{MoS_2}$ have also been widely investigated\cite{t2,t3,t4,t5}, and  the dimensionless  figure of merit, $ZT=S^2\sigma T/(\kappa_e+\kappa_L)$, can measure efficiency of thermoelectric conversion,  in which S, $\sigma$, T, $\kappa_e$ and $\kappa_L$ are the Seebeck coefficient, electrical conductivity, absolute  temperature, the electronic and lattice thermal conductivities, respectively. The strain engineering is a very effective approach to tune  band structure and transport properties in $\mathrm{MoS_2}$\cite{t6,t7,t8}.

Recently, the transition metal
dichalcogenides monolayer $\mathrm{PtSe_2}$  has been  epitaxially grown  with high-quality single-crystal, and the local Rashba spin polarization and spin-layer locking are proved by spin- and angle-resolved photoemission spectroscopy,
which has  potential applications in electrically tunable spintronics and valleytronics\cite{e1,e2}.
Biaxial strain effects on electronic structures and  thermoelectric properties  in  monolayer $\mathrm{PtSe_2}$ have also been
investigated\cite{e3,e4}, and  tensile strain can improve thermoelectric properties by enhancing power factor ($S^2\sigma$) and reducing lattice thermal conductivity ($\kappa_L$). Among  $\mathrm{PtX_2}$  (X=S, Se and Te), $\mathrm{PtTe_2}$ has the smallest energy band gap\cite{e5}, which means small strain may induce semiconductor-metal transition, and has the largest lattice constant $a$\cite{e5}, which implies low lattice thermal conductivity according to tensile strain reduced lattice thermal conductivity in $\mathrm{PtSe_2}$\cite{e4}.
\begin{figure}
  \includegraphics[width=6.0cm]{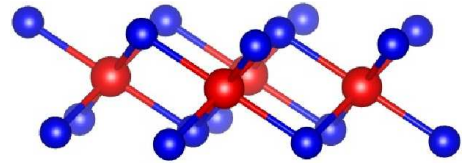}
  \caption{(Color online ) A sketch map of crystal structure of monolayer $\mathrm{PtTe_2}$. The large red balls represent Pt atoms, and small blue balls are Te atoms.}\label{t0}
\end{figure}
\begin{figure*}
  \includegraphics[width=15.0cm]{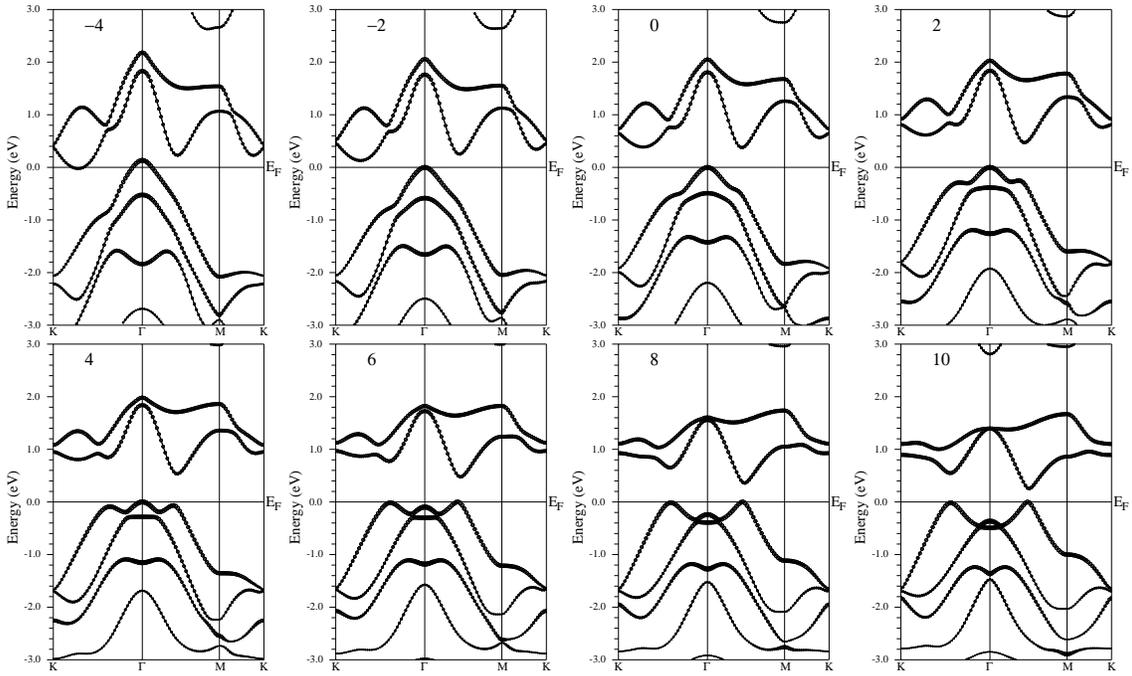}
  \caption{The energy band structures of monolayer $\mathrm{PtTe_2}$ with strain  changing from -4\% to 10\%  using GGA+SOC.}\label{band}
\end{figure*}
\begin{figure}
  \includegraphics[width=8.0cm]{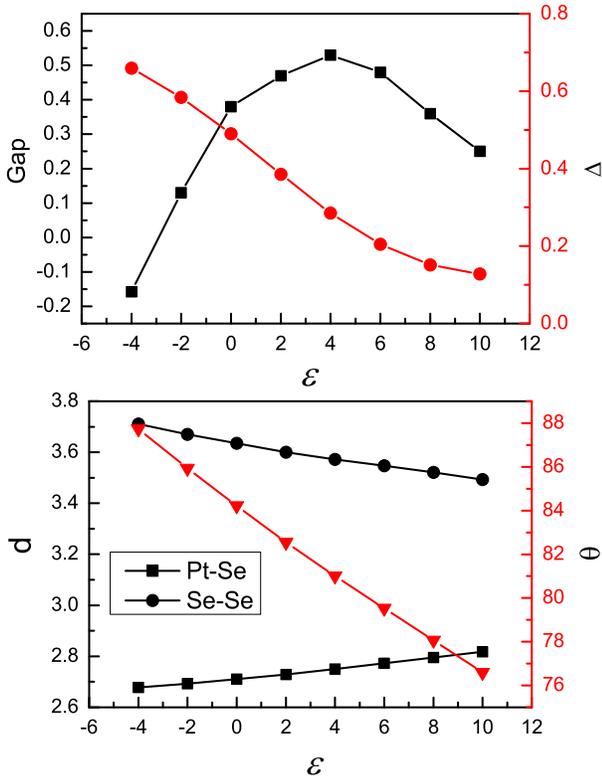}
  \caption{(Color online) The energy band gap Gap (eV),  the value  of spin-orbit splitting at $\Gamma$ point $\Delta$ (eV),
   Pt-Se and Se-Se bond lengths d ($\mathrm{{\AA}}$)  and Se-Pt-Se bond angle $\Theta$  versus biaxial strain $\varepsilon$ using GGA +SOC.}\label{t1}
\end{figure}

Here, we investigate  biaxial strain dependence of electronic structures and  thermoelectric properties of monolayer  $\mathrm{PtTe_2}$  by  first-principle calculations and Boltzmann transport theory.  The SOC has important effects on electronic structures and   power factors in semiconducting transition-metal dichalcogenide monolayers\cite{t8,e4,e5}, so
the electron part is performed  using GGA+SOC, while the lattice part is calculated using GGA.  It is found that energy band gap first increases, and then decreases from compressive strain to tensile strain, which is in excellent agreement
with strain dependence of energy band gap  of other semiconducting transition-metal dichalcogenide monolayers, such as $\mathrm{MoS_2}$\cite{t6,t8} and $\mathrm{PtSe_2}$\cite{e3,e4}.
Calculated results show that the spin-orbit splitting at $\Gamma$ point, Se-Se bond length  and Se-Pt-Se bond angle monotonically decrease, while Pt-Se bond length  monotonically increases. Compressive strain can easily  induce semiconductor-metal transition  as a result of enhanced orbital overlap and hybridization. Tensile strain can induce  valence bands convergence by valley degeneracies, leading to improved Seebeck coefficient, and then can produce enhanced $ZT_e$. Tensile strain can also reduce
lattice thermal conductivity, which has been found in monolayer $\mathrm{PtSe_2}$\cite{e4} and $\mathrm{ZrS_2}$\cite{e4-1}.
Therefore, tensile strain indeed can achieve enhanced thermoelectric properties.

\begin{figure*}
  \includegraphics[width=14cm]{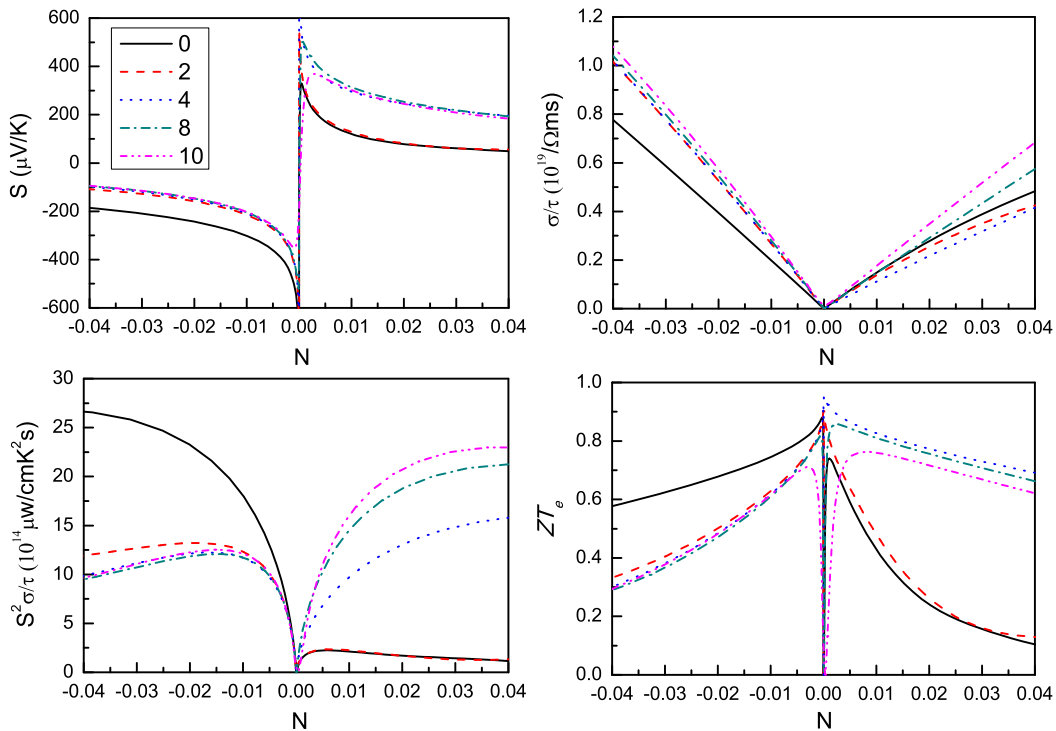}
  \caption{(Color online)  At room temperature,  tensile strain dependence of Seebeck coefficient S, electrical conductivity with respect to scattering time  $\mathrm{\sigma/\tau}$,    power factor with respect to scattering time $\mathrm{S^2\sigma/\tau}$ and  upper limit of $ZT$ $ZT_e$   versus doping level (N) using  GGA+SOC. }\label{t2}
\end{figure*}

The rest of the paper is organized as follows. In the next
section, we shall give our computational details. In the third section, we shall present strain dependence of the electronic structures and  thermoelectric properties of  monolayer  $\mathrm{PtTe_2}$. Finally, we shall give our discussions and conclusions in the fourth section.

\section{Computational detail}
The first-principles calculations are carried out based on
density functional theory\cite{1} as implemented in the  WIEN2k  package\cite{2} within the
full-potential linearized augmented-plane-waves method.  The GGA of Perdew, Burke and  Ernzerhof  (GGA-PBE)\cite{pbe} is used as
the exchange-correlation potential, including all the relativistic  effects for electron part
\cite{10,11,12,so}. The  6000 k-points are used in the
first Brillouin zone (BZ) for the self-consistent calculation. The  free  atomic position parameters  are optimized using GGA-PBE with a force standard of 2 mRy/a.u..
The harmonic expansion up to $\mathrm{l_{max} =10}$ is used in each of the atomic spheres, and $\mathrm{R_{mt}*k_{max} = 8}$ is used to determine  the plane-wave cutoff.  The charge convergence criterion is
used, and  when the integration of the absolute
charge-density difference between the input and output electron
density is less than $0.0001|e|$ per formula unit, the self-consistent calculations are
considered to be converged. Transport calculations
are carried out within the BoltzTrap code\cite{b}, which has been applied successfully to several
materials\cite{b1,b2,b3}.
Since the  accurate transport coefficients need sufficient k-points,  the energy band structures are calculated  using 65 $\times$ 65 $\times$ 11 k-point meshes (50000 k-points) in the
first Brillouin zone.  For 2D materials, the parameter LPFAC usually should choose  larger value. Here, we choose LPFAC value for 20 to achieve the convergence results. Phonon frequencies are obtained using the Phonopy code\cite{at} using 5 $\times$ 5 $\times$ 1 supercell with 8 $\times$ 8 $\times$ 1  Monkhorst-Pack k meshes.
The  lattice thermal conductivities are calculated within  the linearized phonon Boltzmann equation,  which can be achieved by using Phono3py+VASP codes\cite{pv1,pv2,pv3,pv4}. The second order harmonic and third
order anharmonic interatomic force constants  are
calculated by using a  4 $\times$ 4 $\times$ 1  supercell  and a  3 $\times$ 3 $\times$ 1 supercell, respectively. To compute lattice thermal conductivities, the
reciprocal spaces of the primitive cells  are sampled using the 30 $\times$ 30 $\times$ 3 meshes.

\section{MAIN CALCULATED RESULTS AND ANALYSIS}
The monolayer $\mathrm{PtTe_2}$ is composed of  three atomic sublayers with Pt sublayer sandwiched in two Te sublayers,
and  the schematic crystal structure is shown in \autoref{t0}.
The unit cell  of  monolayer $\mathrm{PtTe_2}$  used in the calculations, containing two Te and one Pt atoms,  is constructed with the vacuum region of larger than 15 $\mathrm{{\AA}}$ to avoid spurious interaction, and the optimized lattice constant is $a$=4.02 $\mathrm{{\AA}}$ within GGA-PBE. It has been proved that SOC has important effects on both electronic structures and thermoelectric properties of semiconducting transition-metal dichalcogenide monolayers\cite{t8,e4,e5}.  Therefore, the GGA+SOC is employed to investigate strain dependence of  electronic structures and thermoelectric properties in monolayer $\mathrm{PtTe_2}$.
The unstrained $\mathrm{PtTe_2}$ is a indirect gap semiconductor of 0.38 eV  gap value, with VBM  at $\Gamma$ point and CBM between $\Gamma$ and M points, and the corresponding energy band structure is shown in \autoref{band}.

Both theoretically and experimentally,  strain effects on the electronic structures  of semiconducting transition-metal dichalcogenide monolayers have  been widely studied\cite{t2,t6,t8,e3,e4,e4-1}, some of which  are predicted to produce
semiconductor-metal phase transition by applying suitable strain.
The strain can be simulated by defining  $\varepsilon=(a-a_0)/a_0$, where $a_0$ is the unstrain lattice constant optimized by GGA-PBE, with $\varepsilon$$<$0 ($>$0) being compressive (tensile) strain.
The energy band structures of monolayer $\mathrm{PtTe_2}$ with strain  changing from -4\% to 10\% are plotted in \autoref{band} using GGA+SOC.  The energy band gap,   spin-orbit splitting  value at $\Gamma$ point between the first and second valence bands,
 Pt-Se (Se-Se) bond length and Se-Pt-Se bond angle   versus biaxial strain $\varepsilon$ using GGA +SOC are shown in \autoref{t1}. With strain changing from compressive one to tensile one, the energy band gap  firstly increases, and then decreases, which has been found in many semiconducting transition-metal dichalcogenide monolayers, such as $\mathrm{MoS_2}$\cite{t6,t8}, $\mathrm{ZrS_2}$\cite{e4-1}  and $\mathrm{PtSe_2}$\cite{e4}. It is noteworthy that small compressive strain (about -3\%) can induce semiconductor-metal phase transition, which is less than critical strain of semiconductor-metal phase transition of other 2D materials\cite{t6,e3,e4-1}. As the tensile strain increases, the CBM  moves toward lower energy, while the VBM changes  from the $\Gamma$ point to the one
along the $\Gamma$-M  direction. When the VBM  appears at one point
along the $\Gamma$-M  direction,  the  VBM and  valence band extrema (VBE)  along  $\Gamma$-K  direction are almost degenerate, which can give rise to important effect on Seebeck coefficient. With the increasing compressive strain, the CBM moves from the one point along the $\Gamma$-M  direction to the one along the $\Gamma$-K direction. When the strain reaches  about -3\%, both VBM and CBM cross the Fermi level, inducing semiconductor-metal phase transition, which is due to enhanced orbital overlap and hybridization by reduced Pt-Se bond length.
For  $\mathrm{MoS_2}$, the band gap is sensitively dependent on the S-S or Mo-S bond lengths and S-Mo-S bond angle\cite{jpcl}, the S-S bond length and S-Mo-S bond angle of which  monotonously decrease with strain ($\leq$$\pm$10\%) increasing, and the Mo-S bond length of which monotonously increases.  Similarly,  Se-Se  bond length and Se-Pt-Se bond angle monotonously decrease with increasing  strain considered in the calculations, and Pt-Se bond length monotonously increases.
 When the strain increases from -4\% to 10\% , the spin-orbit splitting at $\Gamma$ point monotonically decreases from 0.66 eV to 0.13 eV. This trend of spin-orbit splitting versus strain is the same with one of monolayer $\mathrm{PtSe_2}$\cite{e4}, while it is opposite to one of $\mathrm{MoS_2}$\cite{t8}.
\begin{figure}[htp]
  \includegraphics[width=8cm]{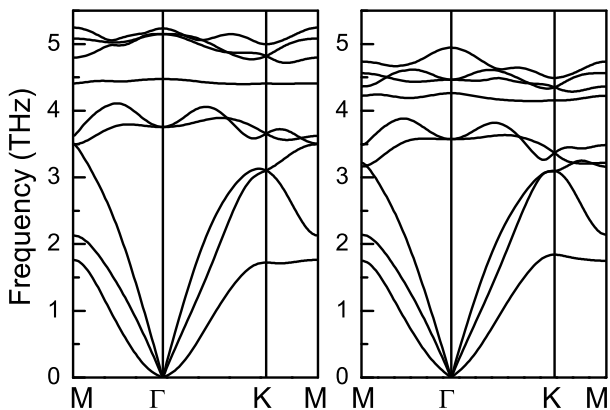}
  \caption{Phonon band structure of $\mathrm{PtTe_2}$ with strain being 0\% (Left) and 4\% (Right) using GGA-PBE.}\label{ph}
\end{figure}

As is well known, bands converge, including orbital  and valley degeneracies, is a very effective strategy to improve Seebeck coefficient, and then to enhance power factor\cite{s1}.  The bands converge can be achieved by strain in many 2D materials such as  $\mathrm{MoS_2}$,  $\mathrm{PtSe_2}$ and   black phosphorus\cite{t8,e4,s2}. According to energy band structures with different strain in \autoref{band}, the  unstrained $\mathrm{PtSe_2}$ nearly has a valley degeneracy of 4 due to the convergence of two  conduction band extrema (CBE) along the $\Gamma$-M  and  $\Gamma$-K directions, which produces largest n-type Seebeck coefficient (absolute value). Tensile strain can lead to the convergence of VBE along the $\Gamma$-M  and  $\Gamma$-K directions, which can lead to enhanced p-type Seebeck coefficient.  Next, the transport coefficients are calculated   based on the constant
scattering time approximation (CSTA) Boltzmann theory within the rigid band approach, being effective for low doping level\cite{tt9,tt10,tt11}. The n-type doping (negative doping level) with the negative Seebeck coefficient can be simulated by shifting the Fermi level into conduction bands, while the p-type doping (positive doping level) with  positive Seebeck coefficient can be achieved by making Fermi level move into valence bands.
\begin{figure}
  \includegraphics[width=8.0cm]{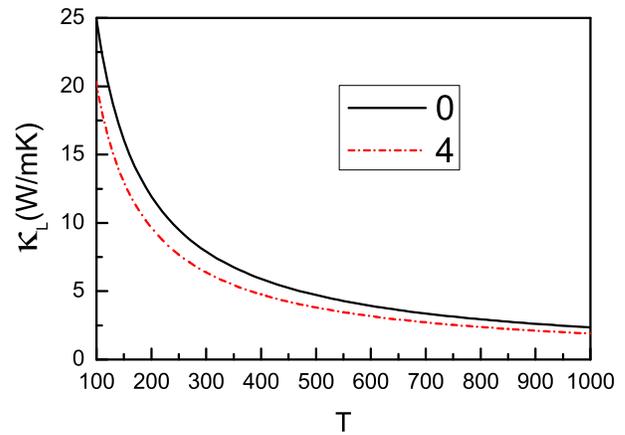}
  \caption{(Color online) The lattice thermal conductivities of $\mathrm{PtTe_2}$ with strain being 0\%  and 4\% using GGA-PBE.}\label{kl}
\end{figure}

To address tensile strain enhanced power factor, the tensile strain dependence of Seebeck coefficient S,  electrical conductivity with respect to scattering time  $\mathrm{\sigma/\tau}$ and  power factor with respect to scattering time $\mathrm{S^2\sigma/\tau}$ using GGA+SOC at room temperature  are plotted in \autoref{t2}. It is clearly seen that the unstrained
$\mathrm{PtTe_2}$ shows the largest n-type Seebeck coefficient, and that tensile strain can induce enhanced Seebeck coefficient.
Calculated results show that the $\mathrm{PtTe_2}$ with strain from 4\% to 10\% shows the nearly same Seebeck coefficient in high doping level. These can be explained by bands converge mentioned above. The complex  strain dependence  of electrical conductivity is observed, due to the sensitive dependence of energy band structures on the applied strain. For power factor combining Seebeck coefficient with electrical conductivity, the unstrained  $\mathrm{PtTe_2}$ has the highest n-type one, while the strain-improved p-type one increases from 4\% to 10\%. An upper limit of $ZT$, taking no account of lattice thermal conductivity,  can be defined as $ZT_e=S^2\sigma T/\kappa_e$, whose tensile strain dependence for $\mathrm{PtTe_2}$ is shown in \autoref{t2}. The highest n-type $ZT_e$ is observed in unstrained   $\mathrm{PtTe_2}$,  while the strain-improved p-type $ZT_e$ in   $\mathrm{PtTe_2}$ with strain from 4\% to 10\% in high doping level is almost the same.

\begin{figure}
  \includegraphics[width=8.0cm]{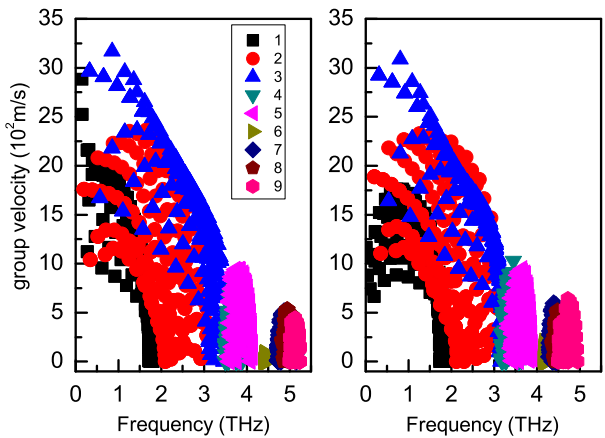}
  \caption{(Color online) Calculated phonon group velocities of $\mathrm{PtTe_2}$ with strain being 0\% (Left) and 4\% (Right)  using GGA-PBE in the irreducible Brillouin zone.}\label{v}
\end{figure}

To further confirm tensile strain-enhanced p-type thermoelectric properties, the lattice  thermal conductivities with strain being 0\% and 4\% are calculated.  Phonon band structures of $\mathrm{PtTe_2}$ with strain being 0\% and 4\% are plotted in \autoref{ph}. It is found that  the longitudinal
acoustic (LA) and transverse acoustic (TA) branches of $\mathrm{PtTe_2}$ with strain being 0\% and 4\% are linear near the $\Gamma$ point, while the z-direction acoustic (ZA) branch  is quadratic near the $\Gamma$ point. The
dispersion of the ZA branch with strain being 4\%  becomes more linear than unstrained one  in the long-wavelength limit.
Calculated results also show that the optical branch with strain being 4\% moves toward lower energy compared to unstrained one.
Thermal conductivity is an intensive property for bulk materials\cite{2dl}.
 The calculated thermal conductivities for 2D materials should be normalized by multiplying Lz/d, where Lz is the length of unit cell along z direction, including vacuum region,  and d is the thickness of 2D materials. However, the thickness of 2D materials is not well defined.  To only prove tensile strain-reduced lattice  thermal conductivity, the unnormalized lattice  thermal conductivities with strain being 0\% and 4\%  are shown in \autoref{kl}. Calculated results show that  the lattice thermal conductivity (6.37 $\mathrm{W m^{-1} K^{-1}}$) at the strain of 4\% can decrease by about 19\% compared to unstrained one (7.89 $\mathrm{W m^{-1} K^{-1}}$) at 300 K. The group velocities in the irreducible BZ with nine different bands are plotted in \autoref{v}. It is found that the group velocities of the first band with strain being 4\%  become more lower than unstrained one, which leads to lower lattice thermal conductivities.
Similar tensile strain-reduced $\kappa_L$ has also  been proved in monolayer $\mathrm{PtSe_2}$\cite{e4} and $\mathrm{ZrS_2}$\cite{e4-1} by the first-principles calculations.
To compare the lattice  thermal conductivities of different 2D materials,  the same thickness d should be used\cite{2dl}. The room-temperature lattice conductivity of $\mathrm{PtTe_2}$  (7.89 $\mathrm{W m^{-1} K^{-1}}$ with Lz being 20.12  $\mathrm{{\AA}}$) is lower than one of $\mathrm{PtSe_2}$ (16.97 $\mathrm{W m^{-1} K^{-1}}$ with Lz being also 20.12  $\mathrm{{\AA}}$) because of larger lattice constants $a$\cite{e4}.  Considering tensile strain-enhanced $ZT_e$ and -reduced $\kappa_L$, tensile strain indeed can improve the efficiency of thermoelectric conversion.

\section{Discussions and Conclusion}
$\mathrm{PtTe_2}$ possesses the smallest gap among the  $\mathrm{PtX_2}$ (X=S, Se and Te) monolayers\cite{e5}, and a semiconductor-to-metal transition can easily be produced for $\mathrm{PtTe_2}$ by strain tuning. Calculated results show that about -3\% compressive strain can induce semiconductor-to-metal transition, which is lower than the critical strain of $\mathrm{MoS_2}$ (about 10\% tensile strain and 15\% compressive strain)\cite{t6} and  $\mathrm{ZrS_2}$ (about 8\% compressive strain)\cite{e4-1}. Experimentally,  such a small strain can easily be achieved by piezoelectric stretching and exploiting the thermal expansion mismatch\cite{bs1,bs2}.

The conduction  bands converge for  unstrained $\mathrm{PtTe_2}$ naturally exists, which is  in favour of better n-type thermoelectric performance (See \autoref{t2}). Both compressive and tensile strain can remove conduction  bands converge, but
tensile strain can induce valence bands converge, which is in support of better p-type  thermoelectric properties. Calculated results show that tensile strain not only can enhance $ZT_e$, but also can reduce $\kappa_L$, which implies that tensile strain
indeed can improve  the efficiency of thermoelectric conversion of $\mathrm{PtTe_2}$ . Similar strain-enhanced thermoelectric properties also can be found in monolayer $\mathrm{MoS_2}$\cite{t8}, $\mathrm{PtSe_2}$\cite{e4} and $\mathrm{ZrS_2}$\cite{e4-1}.

In summary,  we investigate strain dependence of electronic structures and   thermoelectric properties  of monolayer $\mathrm{PtTe_2}$  based mainly on the reliable first-principle calculations and Boltzmann transport theory.
Calculated results show that small compressive strain can give rise to semiconductor-to-metal transition.
It is found that tensile strain can enhance $ZT_e$ and reduce lattice thermal conductivity, and then improve  the efficiency of thermoelectric conversion of $\mathrm{PtTe_2}$.
 So, strain is a very effective method to achieve
tuned electronic and  thermoelectric properties of monolayer $\mathrm{PtTe_2}$, which provides great opportunities for applications in  electronics and thermoelectricity.

\begin{acknowledgments}
This work is supported by the National Natural Science Foundation of China (Grant No. 11404391). We are grateful to the Advanced Analysis and Computation Center of CUMT for the award of CPU hours to accomplish this work.
\end{acknowledgments}

\end{document}